\documentstyle[preprint,aps]{revtex}
\begin{document}
\title{Self-consistent relativistic  calculation of nucleon mean free path}
\author{G. Q. Li\footnote{Present address: Cyclotron Institute,
Texas A\&M University, College Station, TX 83843, USA} and R. Machleidt}
\address{Department of Physics, University of Idaho, Moscow, ID 83843, USA}
\author{Y. Z. Zhuo}
\address{China Institute of Atomic Energy, P. O. Box 275, Beijing, China\\
and Institute of Theoretical Physics, P. O. Box 2735, Beijing, China}
\date{\today}
\maketitle

\begin{abstract}
We present a fully self-consistent and relativistic calculation of
the nucleon mean free path in nuclear
matter and finite nuclei. Starting from the Bonn potential,
the Dirac-Brueckner-Hartree-Fock
results for nuclear matter
are parametrized in terms of an effective
$\sigma$-$\omega$ Lagrangian suitable for the  relativistic density-dependent
Hartree-Fock
(RDHF) approximation. The nucleon mean free path in nuclear matter
is derived from this effective Lagrangian taking diagrams
up to fourth-order into account. For the  nucleon mean free path
in finite nuclei, we make use of the density determined
by the RDHF calculation in the local density approximation.
Our microscopic
results
are in good agreement with the empirical data and predictions by
Dirac phenomenology.
\end{abstract}

\section{Introduction}

One important quantity  of a medium is the mean free path of its
elementary constituents. In nuclear physics, the nucleon mean free path
is of special importance since it can be large compared to the nuclear
size such that the basic assumption of the independent particle motion of
the shell model is reasonable \cite{bohr}.
In nuclear reactions,
the nucleon mean free path is a useful  concept
for summarizing a large number of experimental data \cite{exp1,exp2}.
Furthermore, in the investigation of heavy-ion reactions,
the mean free path of nucleons and other hadrons are often used to
estimate the number of two-body collisions and the reabsorption
effect of the medium on the production cross sections of  hadrons
\cite{li1}.

Most calculations of the nucleon mean free path
have been done in
the framework of non-relativistic dynamics
\cite{schif,coll,nege,mey,fant,blin,li2}, based on e.g. the
phenomenological Skyrme force. Characteristic for the early
theoretical investigations
\cite{schif,coll} is the underestimation of the nucleon mean free path
by up to a factor of two as compared to the empirical value.
The proper treatment of the nonlocality
of the nucleon optical potential resolves much of the discrepancy
between the theoretical prediction and the empirical data \cite{nege,fant}. The
nonlocality of the nucleon optical potential (or mean field) leads to the
reduction of the nucleon mass and consequently  increases
the nucleon mean free path, which is known as the Negele-Yazaki
enhancement \cite{nege}.

Recently there have been some relativistic
calculations of the nucleon mean free
path \cite{cheo,rego,coop,clark},  based on either Dirac phenomenology or the
relativistic impulse approximation for the nucleon optical potential.
In this paper,  we present a fully self-consistent and relativistic
calculation of the nucleon mean free path
in  nuclear matter, starting from the Bonn potential as the realistic
nucleon-nucleon (NN) interaction. The nucleon self-energy (optical
potential) is derived from the Dirac-Brueckner-Hartree-Fock (DBHF)
results for
nuclear matter which includes  the important medium effects.

This work is a continuation of  our effort \cite{mach1,brock1,muth1,li3}
to describe self-consistently
the properties of nuclear matter, finite nuclei and
nuclear reactions based on
the same realistic NN interaction.
There are two aspects to this problem. First, one
needs a realistic NN interaction which is ultimately determined by the
underlying dynamics of quarks and gluons and should in principle be
derived from quantum chromodynamics (QCD). However, due to the
nonperturbative character of QCD in the low-energy regime relevant for
nuclear physics, we are far away from a quantitative understanding
of the NN interaction in this way. On the other hand, there is a good
chance that conventional hadrons, like nucleons and mesons, remain
the relevant degrees of freedom for a wide range of nuclear physics
phenomena. In that case, the overwhelming part of the NN potential
can be constructed in terms of meson-baryon interactions. In fact,
the only quantitative NN interactions available up till now are based
on meson-exchange; a well-known example
is  the
Bonn potential \cite{mach1,mach2} which we apply  in this work.

The second aspect of the problem concerns a suitable many-body
theory  that is able to deal with the bare NN interaction which
has a strong  repulsive core. The Brueckner approach
\cite{brue,beth,gold}
and the variational method \cite{jast,day}
have been developed for this purpose.
However, when using two-body forces,
both many-body theories are not able
to reproduce correctly the saturation properties of
nuclear matter. Inspired by the success of  Dirac phenomenology
in intermediate-energy proton-nucleus scattering \cite{cla1,wall} and
the Walecka model (QHD) for dense nuclear matter \cite{qhd1,qhd2},
a relativistic extension of the Brueckner approach has been initiated
by Shakin and  co-workers \cite{shak1}, frequently called
the Dirac-Brueckner-Hartree-Fock (DBHF)
approach. This approach has been further developed
by Brockmann and Machleidt \cite{mach1,brock1} and by ter Haar and
Malfliet \cite{mal1}.
The common feature of all
DBHF results is that a repulsive relativistic many-body
effect is obtained which is strongly density dependent such that the
empirical nuclear matter saturation can be explained.
The Bonn potential and the
DBHF approach thus provide a reasonable starting point
for  pursuing  the longstanding goal of self-consistently describing
nuclear matter, finite nuclei and nuclear reactions based on
the same realistic NN interaction.

In order to carry out a systematic self-consistent
study of nuclear properties, one usually parametrizes the DBHF
results for nuclear matter in terms of an effective Lagrangian
which, in relativistic density-dependent Hartree Fock (RDHF) approximation,
leads to the same predictions for  nuclear matter
as the original DBHF calculation. The
effective Lagrangian, with its parameters determined by the underlying
NN interaction, can then be used in other domains of nuclear
physics, e.g. the structure of finite nuclei and nuclear reactions.
Different schemes for this parametrization have been proposed
\cite{muth2,marc,brock2,pol}. We use in the present work  the
scheme recently suggested by Brockmann and Toki \cite{brock2}
in which the DBHF results for nuclear matter with the
Bonn potential are parameterized in terms of an effective $\sigma$-$\omega$
Lagrangian. This scheme is however extended to the relativistic
density-dependent Hartree-Fock (RDHF)
approximation, which is more appropriate than
the relativistic density-dependent Hartree (RDH)
approximation of Ref. \cite{brock2}.
The
coupling constants of these effective mesons are  density dependent,
and are determined from the underlying bare NN interaction via a
DBHF calculation.
The nucleon self-energy (optical potential), and hence the nucleon
mean free path,
are then calculated based on
this effective Lagrangian up to the fourth order Feynman diagrams.
For the calculation of the nucleon self-energy and mean free path
in finite nuclei, we use the nucleon density determined by the RDHF
calculation \cite{fritz} in the local density approximation.

We outline the formalism of this work in section II. The results
and discussion are presented in section III. The paper ends with
a brief summary  in section IV.

\section{Formalism}

The relativistic Bonn potential to be used in this work is
constructed in terms of the Thompson equation
which is a three-dimensional
reduction to the original Bethe-Salpeter equation.
The kernel of the Thompson equation, $V({\bf q',q})$, is
the sum of one-meson-exchange amplitudes of certain bosons
with given mass and coupling. In the OBE Bonn model \cite{mach1},
six nonstrange
bosons with mass below 1 GeV are used.
Three sets of  potential parameters, denoted by Bonn A, B
and C, have been proposed (and are given in Table A. 2 of
Ref. \cite{mach1}).
The main difference between the three parameter sets is the cut-off mass
for the $\pi NN$ vertex, which is 1.05, 1.2 and 1.3 GeV for Bonn A,
B and C, respectively. Consequently, the three potentials differ
in the strength of their tensor force component; Bonn A has the
weakest tensor force.
All three  potentials reproduce the deuteron properties and the phase
shifts of NN scattering accurately (cf. Refs. \cite{mach1,brock1}).

The Bonn potential is used in the DBHF calculation for nuclear
matter.
The essential point of this
approach is the use of the Dirac equation
for the description of the single-particle motion in the  nuclear medium
\begin{eqnarray}
[{\bbox \alpha\cdot} {\bf k}+\beta (m+U_S)+U^0_V]
\tilde u({\bf k},s)= E
\tilde u({\bf k},s)\label{eq1}
\end{eqnarray}
where $U_S$ is an attractive scalar field and $U^0_V$ is (the timelike
component of) a repulsive vector field.

As in  conventional Brueckner theory, the basic quantity in the
DBHF approach is the $\tilde G$-matrix which satisfies
the in-medium Thompson equation (also known as relativistic
Bethe-Goldstone equation)
\cite{mach1,brock1,li3}
\begin{eqnarray}
\tilde G({\bf q',q}|{\bf P},\tilde z)=\tilde V({\bf q',q})
+\int {d^3k\over (2\pi )^3}\tilde V({\bf q',k}){\tilde m^2\over
\tilde E^2_{(1/2){\bf P}+{\bf k}}}{Q({\bf k,P})\over 2\tilde E_{(1/2){\bf P}
+{\bf q}}-
2\tilde E_{(1/2){\bf P}+{\bf k}}}\tilde G({\bf k,q}|{\bf P},\tilde z)
\label{eq2}
\end{eqnarray}
with
$$ \tilde m=m+U_S ~{\rm and} ~\tilde E_{\bf k}=({\bf k}^2+\tilde m^2)^{(1/2)}$$
and {\bf P}  the c.m. momentum of the two colliding nucleons in the nuclear
medium.

Since the kernel of the in-medium Thompson equation
depends on the solution of the Dirac equation,
while the Dirac equation  needs  the
scalar and vector potentials which are determined from  the $\tilde G$-matrix,
one is dealing with a self-consistency problem
\cite{mach1,brock1}.
The nuclear matter properties are then obtained from  the in-medium
two-body interaction, the $\tilde G$-matrix. Applying
the Bonn A potential, the
DBHF calculation predicts that nuclear matter saturates at
0.185 fm$^{-3}$ with an  energy per nucleon ${\cal E}/A$=-15.6 MeV,
which is in good agreement with the empirical values.
More results and discussions
concerning the properties of nuclear matter as predicted by the DBHF
approach can be found  in Refs.
\cite{mach1,brock1,li3}.

As proposed by Brockmann and Toki \cite{brock2}, the
DBHF results for nuclear matter can be parametrized by an
effective Lagrangian, in analogy to the $\sigma$-$\omega$ model of
Walecka:
\begin{eqnarray}
{\cal L}=\overline \psi [i\gamma _\mu\partial ^\mu -m-g_\sigma (\rho )
\phi _\sigma -g_\omega (\rho )\gamma _\mu \phi _\omega ^\mu ]\psi
\nonumber
\end{eqnarray}
\begin{eqnarray}
+{1\over 2}(\partial ^\mu \phi _\sigma )^2
{-{1\over 2}m^2_\sigma\phi _\sigma ^2
-{1\over 4}(\partial _\mu \phi _\omega ^\nu -\partial _\nu\phi _\omega ^\mu )^2
+{1\over 2}m^2_\omega \phi _\omega ^{\mu 2}}\label{eq3}
\end{eqnarray}
where $\psi$ is the nucleon field, while
$\phi _\sigma$ and $\phi _\omega ^\mu$
are the effective sigma and
omega fields, respectively. The masses of the effective sigma and omega
mesons are
kept fixed at their values in free-space scattering. However,
the density-dependent coupling constants are choosen such as to reproduce
the DBHF results for nuclear matter
when Eq. (3) is applied in the RDHF approximation.

Treating the effective coupling constants locally as numbers and
calculating in the RDHF approximation, the
nucleon self-energy can be expressed as
\begin{eqnarray}
{\Sigma (k_\mu )=\Sigma _S(k_\mu )+\gamma ^0\Sigma _0(k_\mu )+
{\bbox \gamma \cdot}{\bf k}\Sigma _V(k_\mu )}\label{eq8}
\end{eqnarray}
where $\Sigma _S$, $\Sigma _0$ and $\Sigma _V$ denotes the scalar
component, the time-like part of the vector component and the space-like
part of the vector component of the nucleon self-energy, respectively.
Explicitly the real part of the nucleon self-energy is
given by \cite{qhd2,horo1}
\begin{eqnarray}
\Sigma _S(\rho ,k)=-{2\over \pi ^2}{g_\sigma ^2(\rho )\over
m_\sigma ^2}\int _0^{k_F}dq q^2{m^*_q\over E_q^*}\nonumber
\end{eqnarray}
\begin{eqnarray}
+{1\over 16\pi ^2 k}\int _0^{k_F} dq q {m^*_q\over E^*_q}[g^2_\sigma
(\rho )\Theta _\sigma (k,q)-4g^2_\omega (\rho )\Theta _\omega
(k,q)]
\end{eqnarray}

\begin{eqnarray}
\Sigma _0(\rho ,k)={2\over \pi ^2}{g_\omega ^2(\rho )\over
m_\omega ^2}\int _0^{k_F}dq q^2\nonumber
\end{eqnarray}
\begin{eqnarray}
+{1\over 16\pi ^2 k}\int _0^{k_F} dq q [g^2_\sigma
(\rho )\Theta _\sigma (k,q)+2g^2_\omega (\rho )\Theta _\omega
(k,q)]
\end{eqnarray}

\begin{eqnarray}
\Sigma _V(\rho ,k)=-{1\over 8\pi ^2 k^2}\int _0^{k_F} dq q {q^*\over
E_q^*}[g^2_\sigma
(\rho )\Phi _\sigma (k,q)+2g^2_\omega (\rho )\Phi _\omega
(k,q)]
\end{eqnarray}

where
\begin{eqnarray}
\Theta _i(k,q)={\rm ln}|{A_i(k,q)+2kq\over A_i(k,q)-2kq}|\nonumber
\end{eqnarray}
\begin{eqnarray}
\Phi _i(k,q)={A_i(k,q)\Theta _i (k,q)\over 4kq}-1\nonumber
\end{eqnarray}
\begin{eqnarray}
A_i(k,q)={\bf k}^2+{\bf q}^2+m_i^2-(q_0-k_0)^2,~~~i=\sigma ,\omega\nonumber
\end{eqnarray}
and
\begin{eqnarray}
m_k^*=m+\Sigma _S(\rho ,k)
\end{eqnarray}
\begin{eqnarray}
{\bf k}^*={\bf k}(1+\Sigma _V(\rho ,k))
\end{eqnarray}
\begin{eqnarray}
k_0=({\bf k}^{*2}+m^{*2}_k)^{1/2}+\Sigma _0
\end{eqnarray}

In determining the effective coupling constants from these expressions,
we drop the space-like component of the vector potential and calculate
at the Fermi surface, by identifying $\Sigma _S$ and $\Sigma _0$
with $U_S$ and $U_0^0$ obtained in the DBHF calcaultion.
This is a reasonable assumption since the space-like
component is very small and the potentials are only very weakly momentum
dependent.

There are mainly two  differences
between our effective Lagrangian, Eq. (3),
and the Walecka model \cite{qhd1,qhd2}.
First, the coupling constants in our model are no longer free
parameters fitted to the nuclear matter saturation properties;
these effective coupling constants are  determined
by the DBHF
calculation in which a realistic NN interaction is used.
Second, the coupling constants of our effective
Lagrangian are density dependent, whereas those in the Walecka model are
density independent. The absence of  density dependence in the
Walecka model may be responsible for its
unrealistically large incompressibility.

In the RDHF approximation, the real part of the nucleon self-energy contains
the energy-independent Hartree
contributions (see Fig. 1a)
as well as the energy-dependent Fock contributions (see Fig. 1b).
The lowest order contribution to the imaginary part of the nucleon
self-energy is the fourth-order Feynman diagram which is
characterized by two-particle-one-hole ($2p1h$) intermediate states
(see Fig. 1c). The nucleon
lines in these Feynman diagrams are described by dressed nucleon
propagators, which corresponds to performing the calculation on the
Hartree-Fock ground state and taking account of all Hartree-Fock insertions.
The explicit expressions for the imaginary part of the nucleon self-energy
have been given in Ref. \cite{horo2}.
The derivation of the nucleon self-energy
from the Walecka model has been discussed
in detail in Refs. \cite{horo1,horo2}. For the effective Lagrangian
used in the present work, the expressions for the nucleon self-energy are
the same, but with the coupling constants for sigma- and
omega-exchange replaced by the density-dependent ones as
determined in the nuclear matter  DBHF calculation.

The space-like part of the vector potential, $\Sigma _V$, is rather
small compared to other terms in eq. (\ref{eq8})  and can be absorbed into the
scalar potential and the time-like part of the vector potential by
the following transformation
\begin{eqnarray}
U_S={\Sigma_S-m\Sigma _V\over 1+\Sigma _V},~~U_V={\Sigma _0+E\Sigma _V\over
1+\Sigma _V}\label{eq9}
\end{eqnarray}

In terms of the scalar potential, $U_S$, and the vector potential, $U_V$,
the momentum of a nucleon propagating through a uniform nuclear medium
can be  determined from
\begin{eqnarray}
{E=[(m+U_S)^2+k^2]^{1/2}+U_V}\label{eq10}
\end{eqnarray}
This can be rewritten as
\begin{eqnarray}
{k^2\over 2m}+V+iW=E-m+{(E-m)^2\over 2m}\label{eq11}
\end{eqnarray}
with
$$V=U_{SR}+U_{VR}+{(E-m)\over m}U_{VR}+{1\over 2m}
(U_{SR}^2+U_{VI}^2-U_{SI}^2-U_{VR}^2)$$
$$W=U_{SI}+U_{VI}+{(E-m)\over m}U_{VI}+{1\over m}
(U_{SR}U_{SI}-U_{VR}U_{VI})$$
where we  distinguish between the real and imaginary part
of the scalar and vector potential given by
$$U_S=U_{SR}+iU_{SI}$$
$$U_V=U_{VR}+iU_{VI}$$
Since (V+iW) can be identified as the Schr\"odinger equivalent
potential which is the nucleon optical potential
in the non-relativistic  approach,
Eq. (\ref{eq11}) is identical to the non-relativistic dispersion relation,
except for the  relativistic correction, ${(E-m)^2/2m}$.

Since the potentials $U_S$ and $U_V$ are complex, the nucleon momentum
is also complex and can be expressed as
\begin{eqnarray}
{k=k_R+ik_I}\label{eq12}
\end{eqnarray}
The nucleon mean free path, $\lambda$,
is related to the imaginary part of the nucleon
momentum by \cite{nege}
\begin{eqnarray}
{\lambda ={1\over 2k_I}}\label{eq13}
\end{eqnarray}

 From Eqs. (\ref{eq11}), (\ref{eq12}) and (\ref{eq13})
we obtain an analytical expression for the nucleon
mean free path
\begin{eqnarray}
{\lambda ={1\over 2}\{-m(E-m-V+{(E-m)^2\over 2m})
+m[(E-m-V+{(E-m)^2\over 2m})^2+W^2]^{1/2}\}^{-1/2}}
\label{eq14}
\end{eqnarray}

Since empirical information on the nucleon mean free path
is usually obtained by analysing nucleon-nucleus scattering data, it is
also of interest to perform a microscopic calculation of the  nucleon
mean free path in finite nuclei. To calculate the nucleon self-energy
(optical potential) and mean free path in finite nuclei, one often makes
use of the local density approximation. With this approximation, the
spatial dependence of the nucleon self-energy is directly related to
the density of the nucleus under consideration. Thus in addition to
the expressions for the  nucleon self-energy in nuclear matter, as outlined
above, we also need to know the density of the finite nucleus, so that
we can calculate the  nucleon mean free path in finite nuclei. To attain
self-consistency
of our calculations, the density of the finite nucleus
must be determined in a  RDHF calculation with the effective
Lagrangian, eq. (3). Such a calculation has recently been carried out  by
Fritz {\it et al.} \cite{fritz} for $^{40}$Ca,  which we use in our calculation
of the nucleon mean free path. Note that in finite nuclei, the
incident energy $T_{lab}$ is related to the total energy $E$ by
\begin{eqnarray}
E={m^2+m_T(m+T_{lab})\over ((m+m_T)^2+2m_TT_{lab})^{1/2}}
\end{eqnarray}
where $m$ and $m_T$ are the masses of the nucleon and the  nucleus,
respectively.

\section{Results and discussions}

The coupling constants of the sigma and omega meson
of the effective Lagrangian are shown in Fig. 2a (sigma) and
2b (omega). Both effective coupling constants drop with
increasing density. There are some differences between the
effective coupling constants derived from the Bonn A, B and C
potential; those based on Bonn C  decrease
more with  increasing density. This difference
can be traced back to differences in the tensor-force strength of these
potentials \cite{mach3}.

 From the effective Lagrangian we derive the nucleon
self-energy (optical potential) up to the fourth order
Feynman diagrams (see Fig. 1). In Fig. 3,  we
show the scalar and vector potential, $U_S$ and $U_V$, as
defined by eqs. (\ref{eq8}) and (\ref{eq9}).
We consider two cases with density $\rho$=(1/2)$\rho _0$ (solid
curves) and $\rho _0$ (dashed curves); ~$\rho _0=0.17$ fm$^{-3}$ is
the density of normal nuclear matter. The results are obtained
with the Bonn A potential.
The incident energy $T_{lab}$ is related to the total energy $E$ by
$E=T_{lab}+m$.
Whereas the real part of these potentials depends only weakly on the
incident energy and decreases slightly
(in magnitude) with increasing energy,
the imaginary part depends strongly on the incident energy and
increases (in magnitude) very fast with  energy.

With the scalar potential $U_S$ and the vector potential $U_V$
derived from the effective Lagrangian, we can calculate
self-consistently the nucleon mean free path in nuclear matter
starting from the bare NN interaction. No parameters beyond those
of the Bonn potential are involved in the present calculation.
The nucleon mean free path is calculated from eq. (\ref{eq14}).
The results in normal nuclear matter, which simulates the
interior of heavy nuclei, are shown in Fig. 4. The solid, long-dashed
and short-dashed curves represent the self-consistent results
based on the Bonn A, B and C potential, respectively,
whereas the dotted curve
is the result derived from the original $\sigma$-$\omega$ model of
Walecka (QHD-I)\cite{qhd2}. The dots with error bars represent
early empirical
values of the nucleon mean free path extracted from experiment
\cite{exp3}. The shaded area indicates the estimation of the
nucleon mean free path based on total reaction cross sections
\cite{exp2}. The solid squares correspond to the  recent
empirical data based on Dirac phenomenology in an energy-dependent
analysis \cite{coop,clark}. It is clear that there are still large ambiguities
in the empirical determination of the nucleon mean free path. Within the
error range involved,
our results based on the
Bonn potential are in reasonable agreement with
the empirical data in the whole energy region considered in this
work. Comparing with the
latest empirical data based on  Dirac phenomenology, our results
are slight larger at low energies and in agreement at high energies,
whereas the results based on the QHD-I model are in agreement at low energies
and too small at higher energies.
Between
the self-consistent results corresponding to Bonn A, B and C, there
is also some difference which is due to the differences in the
effective coupling constants (cf. Fig. 2).

Also of interest is the density dependence of the nucleon mean free path.
This information is sometimes needed in the description of heavy-ion
reactions where dense nuclear matter with density up to 2-3$\rho _0$ is
formed. In Fig. 5 we show the density dependence of the nucleon mean free
path,
corresponding to two incident energies. The results are obtained
with the Bonn A potential. The mean free path decreases with the
increasing density, especially at low density.

Finally we shown in Fig. 6 the nucleon mean free path in $^{40}$Ca as a
function
of the radial distance $r$. The solid, dashed and dotted curves
correspond to nucleon energies $T_{lab}$=150, 300 and 450 MeV, respectively.
In the center of the nucleus, the nucleon mean free path is about 2-5 fm,
depending on the energy of nucleon. The nucleon mean free path increases
rapidly at the surface of the nucleus due to the decrease of the nucleon
density.

\section{Summary and outlook}

In this paper,  we have  presented a fully self-consistent calculation
of the nucleon mean free path in nuclear matter and finite nuclei,
starting from a
realistic NN interaction (the Bonn potential). In order to
facilitate a systematic investigation of nuclear properties, we have
parametrized the DBHF results for nuclear matter in terms
of an effective $\sigma$-$\omega$ Lagrangian.
The density
dependent coupling constants
of this effective Lagrangian are determined such as to  reproduce the DBHF
results for nuclear matter in the RDHF approximation. The
effective coupling constants decrease with  increasing density.

In the derivation of the nucleon self-energy (optical potential) based on the
effective Lagrangian, diagrams  up to the fourth order are
included. The
nucleon mean free path is then calculated from the nucleon
optical potential through a  dispersion relation.
Our results for the nucleon mean free path in
nuclear matter are in good agreement
with  empirical data.
We have also analysed the
density and energy dependence of the nucleon mean free path and
find that it decreases with  increasing density and energy.

We have also calculated the nucleon mean free path in finite
nuclei
by means of the local density approximation.
For this purpose,
we made use of the density of $^{40}$Ca as determined in a RDHF calculation
with the effective Lagrangian, eq. (3) \cite{fritz}.
In the center of the nucleus, the nucleon mean free path was found
to be in reasonable agreement with empirical data.

Another aspect is the temperature dependence of the nucleon
mean free path, in addition to its
dependence on  density and energy.
This is of interest for  heavy-ion reactions where
a piece of hot and dense nuclear matter is formed.
This problem is  under investigation.

\vskip 1cm
{\bf Acknowledgement}:  One of the authors (GQL) likes to
thank Prof. H. M\"uther for the RDHF density of $^{40}$Ca.
This work was supported in part by the U.S.
National Science Foundation under Grant No.
PHY-9211607, and by the Idaho State Board of Education.

\pagebreak

\pagebreak

\begin{figure}
\caption{Feynman diagrams for the calculation of the nucleon
self-energy in nuclear matter. (a) Hartree diagram, (b) Fock
diagram, (c) fourth order diagram.}
\end{figure}

\begin{figure}
\caption{Density dependence of the coupling constants for (a) sigma
and (b) omega.}
\end{figure}

\begin{figure}
\caption{Energy and density dependence of the scalar potential
$U_S$ ((a) and (b)) and vector potential $U_V$ ((c) and (d)) in nuclear
matter,
using  the Bonn A potential}
\end{figure}

\begin{figure}
\caption{Energy dependence of the nucleon mean free path in normal
nuclear matter. The solid dots with error bars are from Ref. [40], the shaded
area is from Ref. [3] and the solid squares from Ref. [14]}
\end{figure}

\begin{figure}
\caption{Density dependence of the nucleon mean free path at two
energies. Results are obtained with the Bonn A potential}
\end{figure}

\begin{figure}
\caption{Nucleon mean free path in $^{40}$Ca as function of radial distance
$r$. The solid, dashed and dotted curves correspond to $T_{lab}$=150, 300
and 450 MeV, respectively. The results are obtained with the Bonn A
potential.}
\end{figure}

\newpage

\vspace*{3cm}

The figures
are available upon request from
\begin{center}
{\sc gqli@tamcomp.bitnet}
\end{center}
Please, include your mailing address with your request.


\begin{thebibliography}{99}
\bibitem{bohr}A. Bohr and B. Mottleson, {\it Nuclear Structure},
(Benjamin, New York, 1969).
\bibitem{exp1} R. M. Devries and N. J. DiGiacomo, J. Phys. G {\bf 7}, L51
(1981).
\bibitem{exp2} A. Nadasen {\it et al}., Phys. Rev. C {\bf 23}, 1023 (1981).
\bibitem{li1} G. Q. Li, T. D. Khoa, T. Maruyama, S. W. Huang, N. Ohtsuka,
A. Faessler and J. Aichelin, Nucl. Phys. {\bf A534}, 697 (1991).
\bibitem{schif} J. P. Schiffer, Nucl. Phys. {\bf A335}, 348 (1980).
\bibitem{coll} M. T. Collins and J. J. Griffin, Nucl. Phys. {\bf
A348}, 63 (1980).
\bibitem{nege} J. W. Negele and K. Yazaki,
Phys. Rev. Lett. {\bf 47}, 71 (1981).
\bibitem{mey} H. O. Meyer and P. Schwandt, Phys. Lett. {\bf 107B}, 353 (1981).
\bibitem{fant} S. Fantoni, B. L. Friman and V. R. Pandharipande,
Phys. Lett. {\bf 104B}, 89 (1981).
\bibitem{blin}A. H. Blin, R. W. Hasse B. Hiller and P. Schuck,
Phys. Lett. {\bf 161B}, 211 (1985).
\bibitem{li2} G. Q. Li, J. of Phys. {\bf G17}, 1 (1991).
\bibitem{cheo} T. Cheon, Phys. Rev. C {\bf 38}, 1516 (1988).
\bibitem{rego} R. A. Rego, Phys. Rev. C {\bf 44}, 1944 (1991).
\bibitem{coop} E. D. Cooper, S. Hama, B. C. Clark and R. L. Mercer,
Phys. Rev. C {\bf 47}, 297 (1993).
\bibitem{clark}B. C. Clark, E. D. Cooper, S. Hama, R. W. Finlay
and T. Adami, Phys. Lett. {\bf B 229}, 189 (1993).
\bibitem{mach1} R. Machleidt, Adv. Nucl. Phys. {\bf 19}, 189 (1989).
\bibitem{brock1} R. Brockmann and R. Machleidt, Phys. Rev. C {\bf 42}, 1965
(1990).
\bibitem{muth1} H. M\"uther, R. Machleidt and R. Brockmann, Phys. Rev.
C {\bf 42},  1981 (1990).
\bibitem{li3} G. Q. Li, R. Machleidt and R. Brockmann, Phys. Rev.
C {\bf 45}, 2782 (1992).
\bibitem{mach2} R. Machleidt, K. Holinde and Ch. Elster, Phys. Rep.
{\bf 149}, 1 (1987).
\bibitem{brue} K. A. Brueckner, C. A. Levinson and H. M. Mahmoud,
Phys. Rev. {\bf 95}, 217 (1954).
\bibitem{beth} H. A. Bethe, Phys. Rev. {\bf 103}, 1352 (1956).
\bibitem{gold} J. Goldstone, Proc. R. Soc. (London), {\bf A239}, 267
(1957).
\bibitem{jast} R. Jastrow, Phys. Rev. {\bf 98}, 1479 (1955).
\bibitem{day} V. R. Pandharipande and R. B. Wiringa, Rev. Mod. Phys.
{\bf 51}, 821 (1979).
\bibitem{cla1} L. G. Arnold, B. C. Clark, R. L. Mercer and P. Shwandt, Phys.
Rev. C {\bf 19},  917 (1979).
\bibitem{wall} S. J. Wallace, Ann. Rev. Nucl. Part. Sci. {\bf 37}, 267 (1987).
\bibitem{qhd1} J. D. Walecka, Ann. Phys. {\bf 83}, 491 (1974).
\bibitem{qhd2} B. D. Serot and J. D. Walecka, Adv. Nucl. Phys.
{\bf 16}, 1 (1986).
\bibitem{shak1}M. R. Ansatasio, L. S. Celenza, W. S. Pong and
C. M. Shakin, Phys. Rep. {\bf 100}, 327 (1983).
\bibitem{mal1} B. ter Haar and R. Malfliet, Phys. Rep. {\bf 149}, 207 (1987).
\bibitem{muth2} H. Elsenhans, H. M\"uther and R. Machleidt, Nucl. Phys.
{\bf A515}, 715 (1990).
\bibitem{marc} S. Marcos, M. Lopez-Quelle and N. Van Giai, Phys. Lett.
{\bf B257}, 5 (1991).
\bibitem{brock2} R. Brockmann and H. Toki, Phys. Rev. Lett.
{\bf 68}, 340 (1992).
\bibitem{pol} S. Gmuca, Nucl. Phys. {\bf A547}, 447 (1992).
\bibitem{fritz} R. Fritz, H. M\"uther and R. Machleidt, Phys. Rev. Lett.,
in press.
\bibitem{horo1}C. J. Horowitz and B. D. Serot, Nucl. Phys. {\bf A399}, 529
(1983).
\bibitem{horo2}C. J. Horowitz, Nucl. Phys. {\bf A412}, 228 (1984).
\bibitem{mach3} R. Machleidt and G. Q. Li, Invited talk presented at
"Realistic Nuclear Structure", a conference to mark the 60th brithday of
T. T. S. Kuo, Phys. Rep., in press.
\bibitem{exp3} P. U. Renberg, D. F. Measday, P. Pepin, P. Schwaller,
B. Favier and C. Richard-Serre, Nucl. Phys. {\bf A183}, 81 (1972).
\end{thebibliography}
\end{document}